\newif
\ifconfver
\confvertrue        
\ifconfver
    \documentclass[10pt,conference,letterpaper,twoside]{ieeeconf}
\else
    \documentclass[11pt,draftcls,onecolumn]{IEEEtran}
\fi

\usepackage[utf8]{inputenc}
\usepackage{xspace,empheq,fancybox,amsmath,amssymb,amsfonts,graphicx,epstopdf,epsfig,syntonly,times,amsthm} 
\usepackage{subfig}
\usepackage{psfrag,color,bm,array,tabularx}
\usepackage{subcaption}
\usepackage{cite,url,footnote,xspace,syntonly,algorithmic}
\usepackage{verbatim,multirow}
\usepackage[linesnumbered,ruled,vlined]{algorithm2e}
\usepackage[T1]{fontenc}
\usepackage{textcomp}
\usepackage{hyperref}
\usepackage{xcolor}
\usepackage{mathtools}
\usepackage{makecell}
\usepackage{bbm, booktabs, makecell, subcaption, titlesec}
\newcolumntype{M}[1]{>{\centering\arraybackslash}m{#1}}
\newcolumntype{N}{@{}m{0pt}@{}}

\def\BibTeX{{\rm B\kern-.05em{\sc i\kern-.025em b}\kern-.08em
    T\kern-.1667em\lower.7ex\hbox{E}\kern-.125emX}}

\input{mysymbol.sty}
\allowdisplaybreaks

\IEEEoverridecommandlockouts
\title{\LARGE \bf TRASE-NODEs: \underline{Tra}jectory \underline{Se}nsitivity-aware \underline{N}eural \underline{O}rdinary \underline{D}ifferential \underline{E}quation\underline{s} for Efficient Dynamic Modeling}
\author{Fatima Al-Janahi, Min-Seung Ko, and Hao Zhu
\thanks{F. Al-Janahi, Min-Seung Ko, and H. Zhu are with the Chandra Family Department of Electrical and Computer
Engineering, The University of Texas at Austin, 2501 Speedway, Austin,
TX 78712, USA; Emails: \{FAJ, kms4634500, haozhu\}@utexas.edu.}
\thanks{This work has been supported by NSF Grants 2130706 and 2150571.}
}

\begin{document}
\maketitle

\begin{abstract}
Modeling dynamical systems is crucial across the science and engineering fields for accurate prediction, control, and decision-making. Recently, machine learning (ML) approaches, particularly neural ordinary differential equations (NODEs), have emerged as a powerful tool for data-driven modeling of continuous-time dynamics. Nevertheless, standard NODEs require a large number of data samples to remain consistent under varying control inputs, posing challenges to generate sufficient simulated data and ensure the safety of control design.
To address this gap, we propose trajectory-sensitivity-aware (TRASE-)NODEs, which construct an augmented system for both state and sensitivity, enabling simultaneous learning of their dynamics. This formulation allows the adjoint method to update gradients in a memory-efficient manner and ensures that time-invariant control
set-point effects are captured in the learned dynamics.
We evaluate TRASE-NODEs using damped oscillator and inverter-based resources (IBRs). The results show that TRASE-NODEs generalize better from the limited training data, yielding lower prediction errors than standard NODEs for both examples. The proposed framework offers a data-efficient, control-oriented modeling approach suitable for dynamic systems that require accurate trajectory sensitivity prediction.

\end{abstract}

\section{Introduction} \label{sec:intro}



Dynamical system modeling is a fundamental task for predictive analytics and real-time decision making in various scientific fields such as power systems \cite{DynamicsModeling1}, chemical processes \cite{DynamicsModeling2}, and epidemiology \cite{DynamicsModeling3}. Solving ordinary differential equations (ODEs) 
has widely relied on classical numerical integration methods \cite{NumericalMethods}, such as Euler and Runge–Kutta, 
 which require complete information on the governing dynamics. Advanced machine learning techniques have recently been advocated to use a large amount of sample trajectory data to construct an approximate system model, with notable techniques including ResNet \cite{ResNet} and neural ODEs (NODEs) \cite{Node}. However, improving their sample efficiency and ensuring consistency with control objectives remains challenging when applying ML approaches to large-scale, safety-critical dynamical systems.



Among the ML approaches, NODEs \cite{Node} has been widely adopted thanks to its memory-efficient and continuous-time design; see e.g., \cite{NODEs_Applications1,NODEs_Applications2,NODEs_Applications3}.  By replacing the ODE function with a neural network (NN), it 
keeps the format of continuous-time dynamics, as compared to the fixed discretization grids in ResNet or recurrent NN. The continuous form allows to use adaptive numerical solvers to maintain the smoothness of system trajectories. Furthermore, by using the adjoint sensitivity method to implicitly compute the gradients, NODEs achieves memory-efficient training without any need of unrolling the trajectory. 
However, this unique adjoint-based solution makes it nontrivial to incorporate sensitivity information, which has been shown to greatly enhance sample efficiency and accuracy in static system modeling \cite{SensInteg2,SensInteg3}. 

\begin{figure}[t]
    \centering
    \begin{tabular}{cc}
    \includegraphics[width=0.22\textwidth]{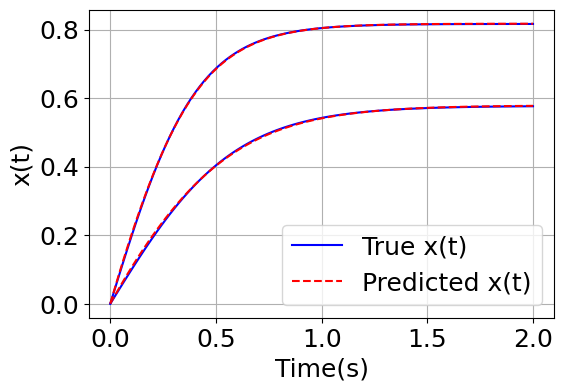} &
    \includegraphics[width=0.22\textwidth]{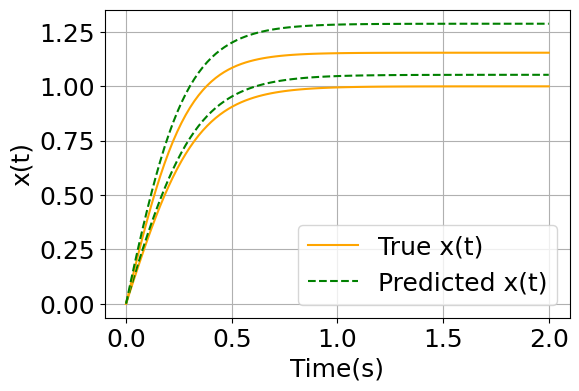} \\
    \small (a) & \small (b)\\
    \end{tabular}
    \centering
    \caption{Comparisons between true and NODE-predicted state trajectories for  a simple linear ODE $\dot x = -3x + u$: (a) training results with both inputs of $u=1,2$ and (b) testing results on unseen inputs of $u=3,4$.}
    \label{fig:fig1}
    \vspace{-0.5cm}
\end{figure}

Meanwhile, sensitivity information is useful for real-world applications because it captures the system’s response to external disturbances or control actions \cite{Sensitivity1}. It plays a central role in system control design and decision making, where understanding the influence of input variations on output trajectories is essential for improving robustness and performance \cite{Sensitivity2}. Yet, standard NODEs focus only on reproducing state trajectories and often overlook this crucial aspect. Even when trained on multiple trajectories with varying inputs, they still fail to capture the correct sensitivity levels or generalize to unseen input patterns.
To demonstrate this issue, we consider a simple example of linear ODE with scalar state and input. Fig.~\ref{fig:fig1} has shown the training and testing results for NODEs with different levels of input set-point.  
Clearly, while NODEs can perfectly match the two training trajectories, there is a significant level of mismatch in predicting the testing trajectories with unseen set-points.  This limitation in generalizability to input set-points arises from NODEs' lack of explicit dependence modeling with a varying input. Incorporating sensitivity into NODEs directly addresses this gap, leading to models that are more robust to unseen or perturbed inputs. This way, it also requires less samples and avoids repeated simulations, thereby reducing the burden on data generation or collection. 

Recent works have extended NODEs to model input-driven dynamical systems by incorporating control variables or external inputs into the learned dynamics \cite{LitReview1, LitReview2,LitReview3}. These approaches demonstrate that NODE-based models can represent controlled dynamical behavior and have been applied to tasks such as system identification, control learning, and enforcing structural properties of physical systems. Although these methods can improve generalization by imposing structure on the learned dynamics, they typically do not explicitly characterize how system states locally depend on input variations. As a result, the resulting models may lack explicit knowledge of input–trajectory sensitivities, which describe how system states respond to small variations in inputs. Incorporating such sensitivity information provides additional insight into the input-state relationship, which can further improve the model’s generalization ability to unseen input conditions.

Our work seeks to develop the trajectory-sensitivity-aware (TRASE-)NODEs, pursuing the sensitivity learning enhancement for the adjoint method. It neatly constructs an augmented ODE system for the state and sensitivity for learning both simultaneously, ensuring that the influence of control input is captured together with the underlying dynamics. Our proposed TRASE-NODEs still rely on the predictive modeling of the original ODE function. However, this augmented formulation makes it possible to update the adjoint analysis for computing the gradient in a memory-efficient fashion. The resultant augmented adjoint method requires to compute second-order partial derivatives. At an increased computation complexity, our model 
 can improve the prediction  accuracy while generalizing more effectively to unseen inputs.
 We investigated the effectiveness of TRASE-NODEs through two case studies. The first involves a damped oscillator, which represents a simple ODE with a known analytical form, while the other considers an inverter-based resource (IBR) system with higher-order nonlinear dynamics. Numerical results demonstrate the improved generalization of TRASE-NODEs, with orders-of-magnitude improvement over NODEs on the damped oscillator. For the practical IBR system, TRASE-NODEs more accurately capture the state sensitivity to the voltage reference, enabling more effective control design for power system stability analysis.


\section{Problem Statement} \label{sec:PS}



We briefly discuss the ODE  and the NODEs approach for formulating our sensitivity-aware design. Consider a continuous-time dynamical system with the state vector denoted by $\bbx(t) \in \mathbb R^n$ for the time $t\geq 0$. In addition, an external input of the reference set-point, denoted by $\bbu \in \mathbb R^k$, is used to control the system. Thus, we consider a generally nonlinear, time-dependent ODE:
\begin{align}
    &\frac{d\bbx}{dt} = \bbf(\bbx(t),t, \bbu),~~\mathrm{with}~~\bbx(0) = \bbx_0. \label{eq:ode}
\end{align}
where the function $\bbf(\cdot)$ is the key to defining the state evolution. Note that our setup includes a time-invariant control set-point $\bbu$, which goes beyond the original NODEs work \cite{Node}. Including $\bbu$ is very important for dynamics control problems with actuation capability, as discussed later. Additional inputs to $\bbf(\cdot)$ could include time-varying algebraic variables collected in vector $\bby(t) \in \mathbb R^m$. They are determined by algebraic relations or external system conditions rather than differential equations. Thus, their trajectories are not obtained through ODE-based integration, but they evolve over time due to their dependence on $\bbx(t)$ and other exogenous inputs. When $\bby(t)$ directly influences the evolution of $\bbx(t)$, it is necessary to include the former in $\bbf(\cdot)$. For example, the dynamics of inverter-based resources, or general power system components, depend on the terminal bus voltage magnitude and angle (frequency), which are the algebraic variables in this case \cite{DynamicsModeling1}. The presence of such algebraic variables does not affect the proposed method, since they can be incorporated in the same framework as additional inputs influencing the state dynamics.



By connecting discrete ODE solvers with ResNet in \cite{ResNetNode1,ResNetNode2}, the NODEs work  \cite{Node} proposed 
a continuous-time neural approximation of $\bbf(\cdot)$. A neural network (NN) with weight/bias parameters in $\bbtheta$ is used to approximate a given time window of $[0,T]$, resulting in another ODE: 
\begin{align}
   \frac{d\bbx}{dt} = \bbf_\bbtheta (\bbx(t), t, \bbu),~\forall t\in[0,T].\label{eq:node}
\end{align}
Upon selecting the NN architecture (e.g., the simple feedforward NN), the problem becomes finding the best $\bbtheta$ that minimizes a chosen loss function for $\{\hhatbbx_\bbtheta(t)\}_{t\in [0,T]}$. The latter represents the predicted trajectory generated by integrating \eqref{eq:node} from the initial $\bbx_0$.
This makes it possible to evaluate the loss function $\mathcal{L}( \hhatbbx_{\bbtheta} (t), \bbarbbx(t))$ with given true trajectory $\{\bbarbbx(t)\}_{t\in [0,T]}$. To train the NODE, one need to compute the gradient with respect to (wrt) $\bbtheta$, namely $\tfrac{\partial \mathcal{L}}{\partial \bbtheta}$.  
Gradient computation for NODEs could suffer from excessive memory costs caused by storing the states and gradients along the full integration path.
To address this issue, the adjoint sensitivity method was used in \cite{Node}, which is a classical technique in optimal control theory to reduce the memory needs for gradient computation \cite{OptimalControlTheory}. Fig.~\ref{fig:NODEs_VS_TRASENODEs} illustrates the forward and backward passes of NODEs training based on adjoint method. 
By defining the adjoint state $\bba(t) = \frac{\partial \mathcal{L}}{\partial \bbx(t)}$, it  evolves according to another ODE: 
\begin{align}
   \frac{d\bba}{dt} = -\big[\bba(t)\big]^\top\frac{\partial \bbf_\bbtheta}{\partial \bbx(t)} .\label{eq:adjode}
\end{align}
By integrating \eqref{eq:adjode} backward in time, the trajectory of $\bba(t)$ can bed obtained and used to solve for the gradient:
\begin{align}
   \frac{\partial \ccalL}{\partial \bbtheta} = -\int_{T}^{0} \big[\bba(t)\big]^\top \frac{\partial \bbf_\bbtheta}{\partial \bbtheta} \,dt\ .\label{eq:nodegradient}
\end{align}
As illustrated in Fig.~\ref{fig:NODEs_VS_TRASENODEs}, the forward pass only needs to integrate \eqref{eq:node} to obtain the predicted the terminal $\bbx(T)$. In the backward pass, integrating \eqref{eq:adjode} starting from $t=T$ allows to form the trajectory of $\bba(t)$, which allows to compute the gradient  $\frac{\partial \ccalL}{\partial \bbtheta}$ using \eqref{eq:nodegradient}. 
Thus, this implementation no longer needs to store the integration path in the forward pass, but only keeps the terminal $\bbx(T)$. This can significantly improve the scalability of NODEs training. 

However, the vanilla NODEs approach is limited to matching the state trajectory only and does not account for variations in ODE parameters such as control set-points. The trajectory sensitivity wrt~$\bbu$ is very important for optimal control design, parameter estimation, and stability analysis, widely used in diverse domains; see e.g., \cite{TRASE_Applications}.
For example, it is a powerful power system dynamic analysis tool used to guide device placement, design controller parameters, and efficiently optimize corrective actions for decision support \cite{Trase_in_PS1, Trase_in_PS2, Trase_in_PS3}. 
For simplicity, our work focuses on its formulation for a scalar $\bbu$, but it is readily extended to vector $\bbu$ by using concatenation. The trajectory sensitivity wrt~$\bbu$ is defined as 
\begin{align}
   \bbs(t) = \frac{\partial \bbx(t)}{\partial \bbu} \Big |_{\bbx(0) = \bbx_0}, \label{eq:sensode}
\end{align}
which has the same dimension as $\bbx(t)$. Basically, it captures the impact of a small perturbation in $\bbu$ on the state trajectory. 
When $\bbs(t)$ is not explicitly considered, the resultant NODEs could be much less effective in generalizing to unseen control input scenarios. While we could mitigate this issue by using an enlarged or denser space for sampling the control set-points, this approach inevitably increases data requirements and training costs. Thus, we put forth a new NODEs design that directly incorporates $\bbs(t)$ into the training process, to improve data efficiency and generalizability to different control  actions \cite{SensitivityGeneralization}.

\section{TRASE-NODEs via Adjoint Method} \label{sec:TRASE}


\begin{figure}[t]
\centering
\smallskip 
\includegraphics[scale=0.21]{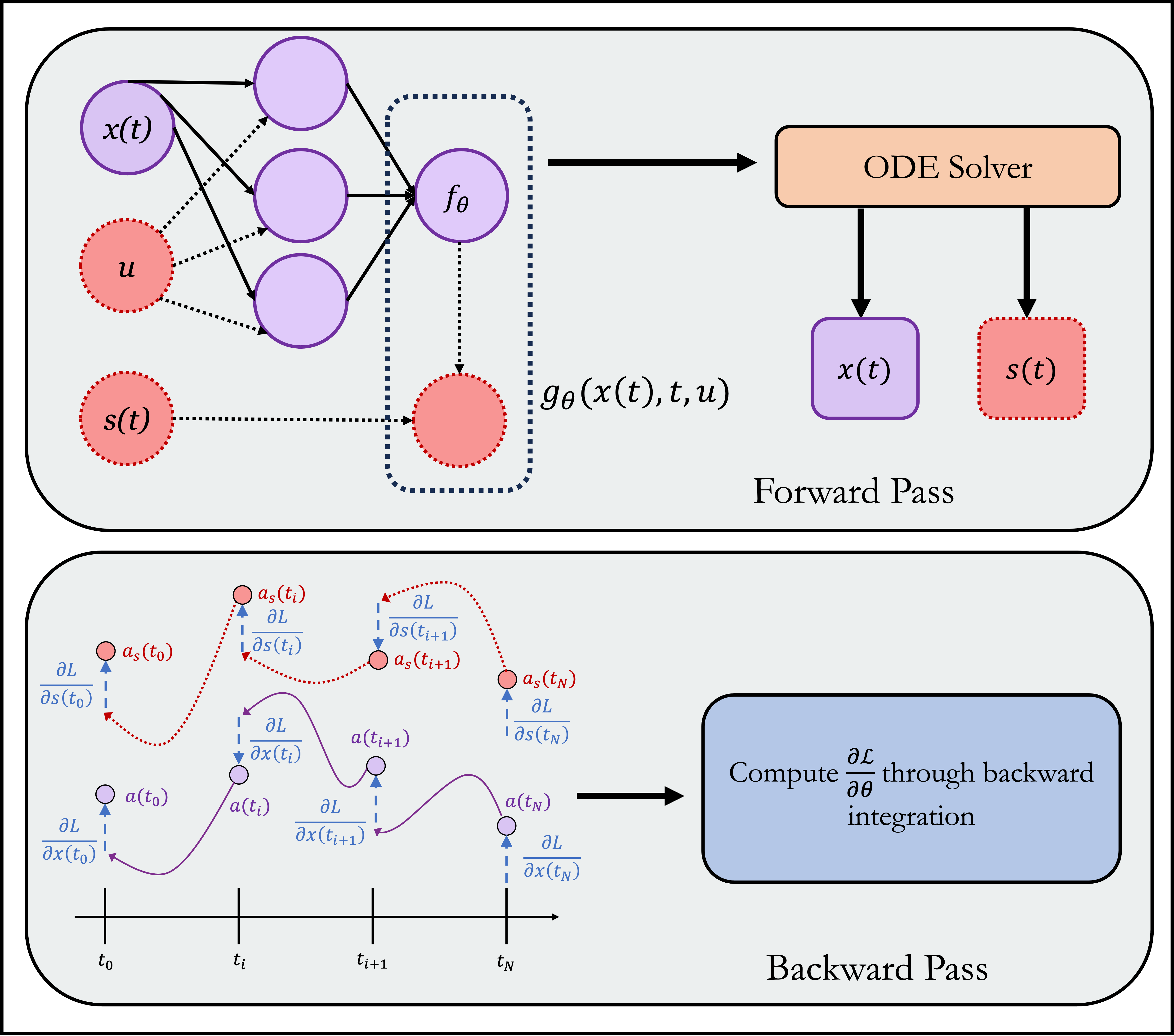}
\caption{Proposed TRASE-NODEs training in dotted lines builds upon updating the NODEs framework in solid lines.}
\label{fig:NODEs_VS_TRASENODEs}
\vspace{-0.5cm}
\end{figure}

We present the trajectory-sensitivity-aware (TRASE-)NODEs framework implemented using the adjoint method. To this end, we use the original NN $\bbf_\bbtheta(\cdot)$ for the NODEs modeling but  augment the ODE in \eqref{eq:node} to also include the time-varying  trajectory sensitivity. 
Fig.~\ref{fig:NODEs_VS_TRASENODEs} includes the  updates of the proposed TRASE-NODEs design, with $\bbs(t)$ being part of the augmented state. 
Specifically, the augmented state vector $\bbz(t) = [\bbx(t);\bbs(t)]\in \mathbb R^{2n}$ evolves per the augmented ODE as 
\begin{align}
    \frac{d\bbz}{dt} = \bbg_\bbtheta(\bbz(t),t,\bbu) =  \begin{bmatrix}
    \bbf_\bbtheta\\[5pt]
    \displaystyle \frac{\partial  \bbf_{\bbtheta}}{\partial \bbu}+\frac{\partial \bbf_\bbtheta}{\partial \bbx(t)}\times \bbs(t)
    \end{bmatrix},  \label{eq:augstate}
\end{align}
where the chain rule is used here to form the time-differentiation for $\bbs(t)$. This augmented function $\bbg_\bbtheta(\cdot)$ uses both $\bbx(t)$ and $\bbs(t)$ as input, but is parameterized by the same vector $\bbtheta$ as in $\bbf_\bbtheta(\cdot)$. The augmented state $\bbz(t)$ remains fully differentiable wrt $\bbtheta$, making gradient-based training possible.
More importantly, this update allows to use the loss function $\mathcal{L}( \hhatbbz_{\bbtheta} (t), \bbarbbz(t))$ which can match not only the state trajectory $\bbx(t)$, but also its sensitivity $\bbs(t)$. Note that the same forward-pass procedure as the original NODEs can be applied to $\bbg_\bbtheta(\cdot)$ here. As shown in Fig.~\ref{fig:NODEs_VS_TRASENODEs}, the forward pass can easily incorporate $\bbs(t)$ and $\bbg_\bbtheta(\cdot)$ to solve for $\bbz(t)$. But the adjoint method in the backward pass requires some major changes to account for the inclusion of the sensitivity state $\bbs(t)$.




Let us define the augmented adjoint vector $\bba^\bbz(t)$ for the full state $\bbz(t)$, given by 
\begin{align}
   \bba^\bbz(t): =\frac{\partial \ccalL}{\partial \bbz(t)}=\begin{bmatrix}
    \frac{\partial \ccalL}{\partial \bbx(t)} \\
    \frac{\partial \ccalL}{\partial \bbs(t)}
    \end{bmatrix} =\begin{bmatrix}
    \bba^\bbx(t)\\
    \bba^\bbs(t)
    \end{bmatrix}. \label{eq:augadj}
\end{align}
\begin{proposition}[Augmented adjoint ODE] \label{prop:aug} The augmented $\bba^\bbz(t)$
 follows the ODE [cf.~\eqref{eq:adjode}]:
\begin{align}
   \frac{d \bba^\bbz(t)}{dt}=- \big[\bba^\bbz(t)\big]^\top \frac{\partial \bbg_\bbtheta
   }{\partial \bbz(t)} \label{eq:augadjode}
\end{align}
with the Jacobian matrix given by
\begin{align}
    \frac{\partial \bbg_{\bbtheta}
    }{\partial \bbz(t)}
    = \begin{bmatrix}
    \displaystyle \frac{\partial \bbf_{\bbtheta}}{\partial \bbx(t)} & {\mathbf{0}} \\[8pt]
    \displaystyle \frac{\partial^2 \bbf_{\bbtheta}}{\partial \bbx(t)\,\partial \bbu}
    + \frac{\partial^2 \bbf_{\bbtheta}}{\partial \bbx(t)^2}\times\bbs(t)
    & \displaystyle \frac{\partial \bbf_{\bbtheta}}{\partial \bbx(t)}
    \end{bmatrix}. \label{eq:dfdy}
\end{align}
\end{proposition}

Proposition \ref{prop:aug} shows that although the  ODE structure is the same as \eqref{eq:adjode},  
the corresponding Jacobian in~\eqref{eq:dfdy} is more complicated due to the second-order partial derivatives of $\bbf_\bbtheta(\cdot)$. For the latter, its first diagonal block is exactly the Jacobian matrix in \eqref{eq:adjode}, because the dynamics for $\bba^\bbx(t)$ remains the same. The inclusion of second-order derivatives or Hessian matrices would increase the computational cost of the proposed TRASE-NODEs, as detailed soon. 
But still, the same training procedure follows from NODEs by integrating \eqref{eq:augadjode} backward in time starting from the terminal $\bba^\bbz(T)$, 
which can be easily obtained from automatic differentiation of the loss at the terminal time $t=T$.

\begin{proposition}[Adjoint-based gradient computation] \label{prop:grad}

Upon obtaining the $\bba^\bbz(t)$ trajectory, the backward pass can compute the gradient using
\begin{align}
   \frac{\partial \mathcal{L}}{\partial \bbtheta} = -\int_{T}^{0}\big[\bba^\bbz(t)\big]^\top \frac{\partial \bbg_\bbtheta}{\partial \bbtheta} \,dt\ \label{eq:trasenodegrad}
\end{align}
with the augmented Jacobian matrix given by
\begin{align}
   \frac{\partial \bbg_\bbtheta}{\partial \bbtheta}=\begin{bmatrix}
    \displaystyle \frac{\partial \bbf_\bbtheta}{\partial \bbtheta} \\[5pt]
    \displaystyle \frac{\partial^2 \bbf_\bbtheta}{\partial \bbtheta \partial \bbu}+\frac{\partial^2 \bbf_\bbtheta}{\partial \bbtheta \partial \bbx(t)}\times \bbs(t)
    \end{bmatrix}.\label{eq:dfdtheta}
\end{align}
\end{proposition}

Proposition \ref{prop:grad} shows that forming the augmented Jacobian matrix has again increased the complexity due to the 
second-order partial derivatives. To compute these terms for both \eqref{eq:dfdy} and \eqref{eq:dfdtheta}, we adopt a two-stage procedure. We first compute the NODEs-induced Jacobian matrices $\displaystyle \frac{\partial \bbf_\bbtheta}{\partial \bbtheta}$ and $\displaystyle \frac{\partial \bbf_\bbtheta}{\partial \bbx(t)}$  via automatic differentiation. Second, for each entry of these two matrices, we further use automatic differentiation to evaluate its derivative wrt a chosen variable (e.g., state variable, set-point $\bbu$, or NN parameter), thereby constructing the corresponding Hessian matrix. 
The second stage could significantly increase the computation complexity depending on the dimensions of state $\bbx(t)$ and 
control set-point. But the memory efficiency is maintained for the TRASE-NODEs training thanks to the adjoint method. 

To provide more details on the training of TRASE-NODEs, it follows the original NODEs but accounts for the new state $\bbs(t)$, as illustrated in Fig.~\ref{fig:NODEs_VS_TRASENODEs}. The forward pass integrates \eqref{eq:augstate} to obtain the final state $\bbz(T)$. Using the latter, the backward pass can form the terminal $\bba^\bbz(T)$ and use it to integrate \eqref{eq:augadjode} in reverse time to generate the trajectory of $\bba^\bbz(t)$.  To detail this integration process, consider a total of $N$ time steps within $[0,T]$,
denoted by $\{t_i\}_{i=1}^N$. At each integration step $i$, the inputs are the current augmented state $\bbz(t_i)$ and the adjoint state $\bba_{\bbz}(t_i)$, used to obtain both $\bbz(t_{i-1})$ by integrating~\eqref{eq:augstate}
backward, and the adjoint state $\bba_{\bbz}(t_{i-1})$ by using~\eqref{eq:augadjode}. Accordingly, the accumulated gradient till step $i$, namely $\frac{\partial \mathcal{L}}{\partial \bbtheta}(t_{i})$, can incorporate the contribution from step $i$ to form $\frac{\partial \mathcal{L}}{\partial \bbtheta}(t_{i-1})$. 
This backward process continues recursively until reaching the initial time $t_1=0$, allowing efficient backpropagation through the augmented adjoint system without storing the entire forward trajectory.

\section{Simulation Results} \label{sec:results}


We present the numerical evaluation results for  the proposed TRASE-NODEs by using two test systems. The first test uses a damped oscillator to illustrate performance on a simple, well-understood dynamic system. The second one considers an IBR, showing the applicability to more complex system with high-order dynamics and algebraic variables.
In both cases, the augmented dynamics are approximated using a fully connected feed-forward NN. Sensitivity trajectories are obtained from the learned dynamics $\bbf_\bbtheta$ via \eqref{eq:augstate}, and the parameters are trained using mean-squared error (MSE) loss over both state and sensitivity trajectories with the $Adam$ optimizer.

\vspace{3pt}

\noindent \textbf{Test Case 1: Damped Oscillator.} 
This is a classical second-order dynamical system, given by
\begin{align}
    \ddot{x}(t) + 2 \zeta \omega_n \dot{x}(t) + \omega_n^2 x(t) = u
\end{align}
where $x(t)$ is the displacement with the velocity $v(t)=\dot{x}(t)$ driven by an external forcing input $u$. The system parameters include the natural frequency $\omega_n$ and  damping ratio $\zeta$.  Its dynamics typically exhibits oscillatory motion with an amplitude that decays over time due to the damping effect. We can convert it to a first-order system with the state vector $[x(t); v(t)]$, and analytically obtain the ODE for its corresponding sensitivity state using \eqref{eq:augstate}, as given by  
\begin{align}
    \dot{s}_x(t) &= s_v(t), \\
    \dot{s}_v(t) &= - (\omega_n^2) s_x(t) -(2\zeta\omega_n) s_v(t)  + 1.    \label{eq:sxv}
\end{align}
Hence, the augmented system has four state variables in $\bbz(t)$. 

For modeling $\bbf_\bbtheta$, we use a feedforward NN that has  a single hidden layer with 32 units and \textit{LeakyReLU} activation, followed by a linear output layer. The NN input consists of  $[x(t);v(t)]$ and the scalar set-point $u$.
To generate the training trajectory, 
we use the set-point $u=5$, and set the parameters as $\omega_n = 2.5$ and $\zeta=0.3$.
Using a time window of 
$[1,7]$, we implement an ODE solver with the initial conditions of $x(0)=2$, $v(0)=1$, and $s_x(0)=s_v(0)=0$. The resultant augmented trajectory, sampled using 100 uniformly spaced time-steps, serves as the ground-truth reference for prediction. 
For training TRASE-NODEs, only a single scenario with $u=5$ of both state and sensitivity trajectories were used. In contrast, NODEs model was trained by using two scenarios of state trajectories  with $u=5$ and $u=5.1$. 

\begin{figure}[t]
    \centering
    \begin{tabular}{cc}
    \includegraphics[width=0.22\textwidth]{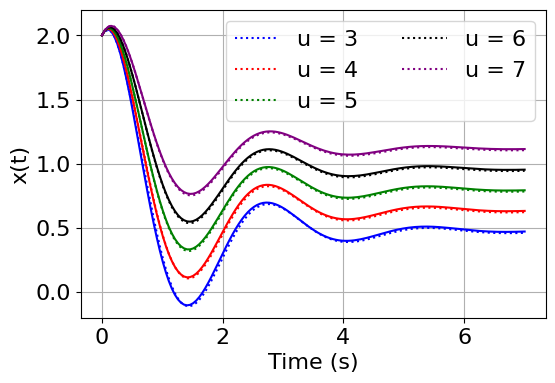} \label{fig:a} &
    \includegraphics[width=0.22\textwidth]{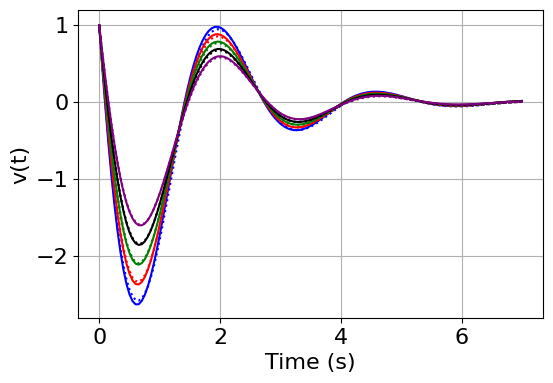} \label{fig:b} \\
    \small (a) & \small (b)\\
    \includegraphics[width=0.22\textwidth]{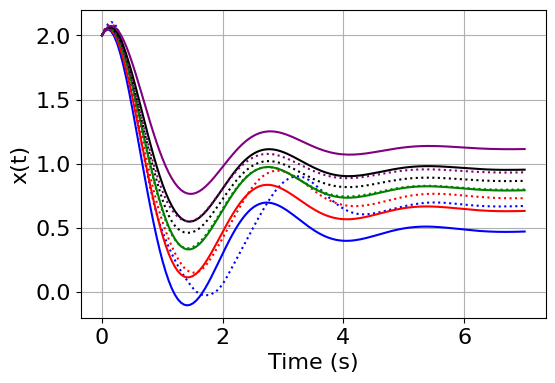} \label{fig:c} &
    \includegraphics[width=0.22\textwidth]{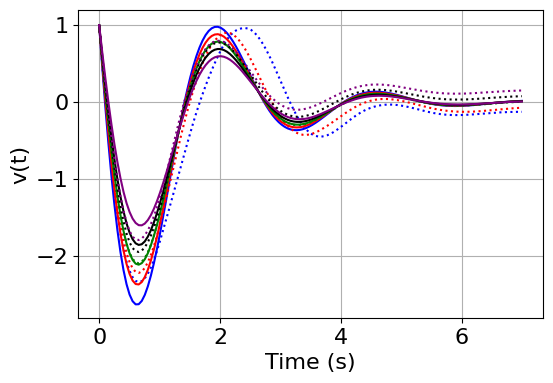} \label{fig:d} \\
    \small (c) & \small (d)\\
    \end{tabular}
  \caption{Comparison of state trajectories with different $u$ values: (a) $x$ and (b) $v$ for TRASE-NODEs and (c) $x$ and (d) $v$ for NODEs. Solid lines denote ground-truth trajectories, with dotted lines for predictions.}
  \label{fig:Ex1_x(t)}
  \vspace{-0.4cm}
\end{figure}

\begin{figure}[t]
    \centering
    \begin{tabular}{cc}
    \includegraphics[width=0.22\textwidth]{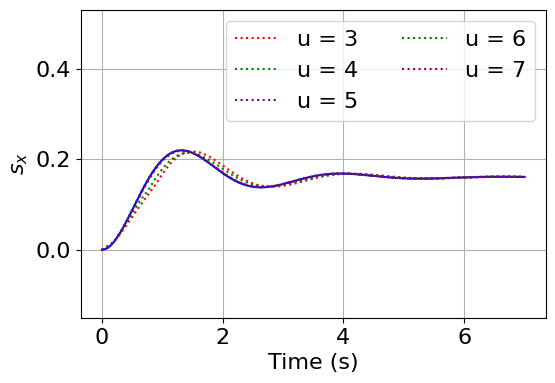} \label{fig:a} &
    \includegraphics[width=0.22\textwidth]{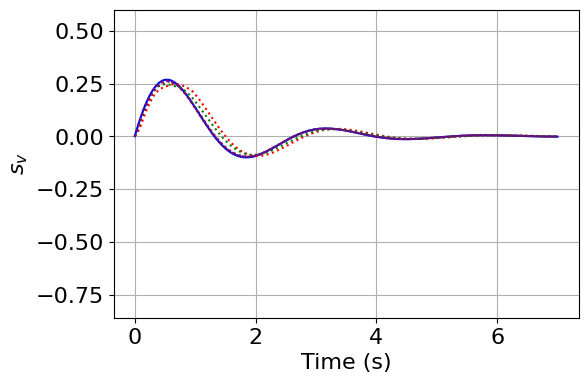} \label{fig:b} \\
    \small (a) & \small (b)\\
    \includegraphics[width=0.22\textwidth]{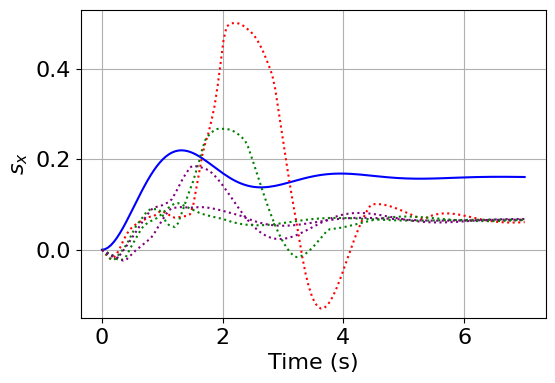} \label{fig:c} &
    \includegraphics[width=0.22\textwidth]{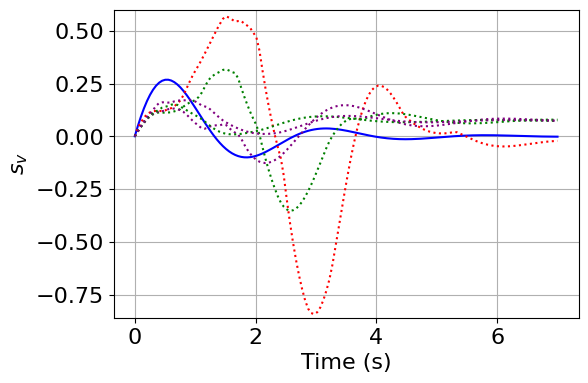} \label{fig:d} \\
    \small (c) & \small (d)\\
    \end{tabular}
     \caption{Comparison of state trajectories with different $u$ values: (a) $s_x$ and (b) $s_v$ for TRASE-NODEs and (c) $s_{x}$ and (d) $s_{v}$ for NODEs. Solid lines denote ground-truth trajectories, with dotted lines for predictions.} 
  \label{fig:Ex1_s(t)}
  \vspace{-0.3cm}
\end{figure}

\begin{figure}[t]
    \centering
    \begin{tabular}{cc}
    \includegraphics[width=0.22\textwidth]{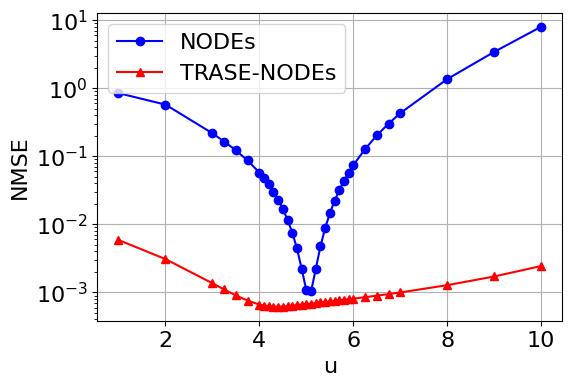} \label{fig:a} &
    \includegraphics[width=0.22\textwidth]{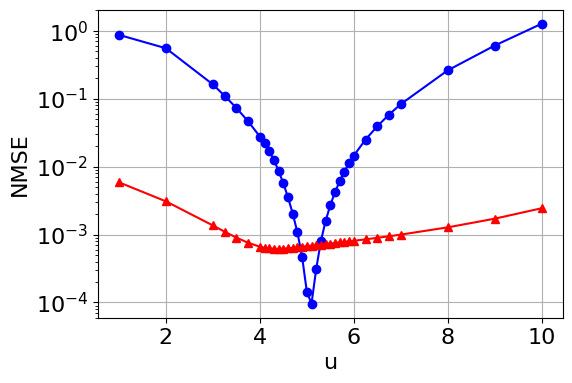} \label{fig:b} \\
    \small (a) & \small (b)\\
    \includegraphics[width=0.22\textwidth]{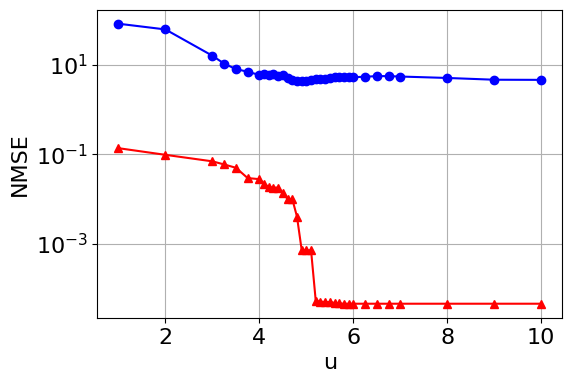} \label{fig:c} &
    \includegraphics[width=0.22\textwidth]{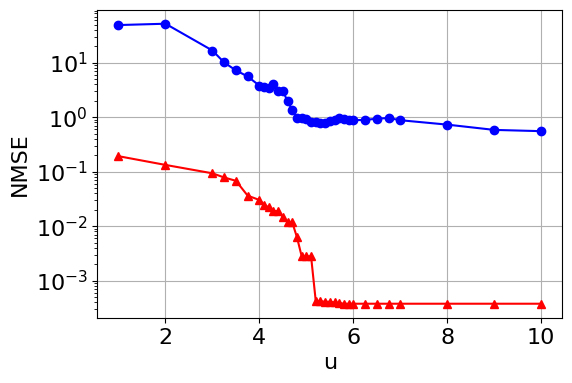} \label{fig:d} \\
    \small (c) & \small (d)\\
    \end{tabular}

  \caption{Comparison of NMSE for TRASE-NODEs and NODEs in logarithmic scale over the range of input $u$ for (a) $x$,  (b) $v$, (c) $s_x$, and (d) $s_v$.} 
  \label{fig:Ex1_NMSE}
  \vspace{-0.5cm}
\end{figure}

We compare TRASE-NODEs with the baseline NODEs under perturbed control inputs of $u\in[1,10]$. Representative trajectories for both states and sensitivities are shown in Fig.~\ref{fig:Ex1_x(t)} and Fig.~\ref{fig:Ex1_s(t)}, respectively. In addition,  Fig.~\ref{fig:Ex1_NMSE} plots the overall normalized MSE (NMSE) error over the full range of $u$ values. Clearly, the proposed TRASE-NODEs achieves consistently lower error across all testing cases. The worst-case NMSE values for predicting $\bbz(t) = [x; v; s_x; s_v]$ are $[0.006; 0.0094; 0.14; 0.19]$ for TRASE-NODEs, as  compared to $[7.9; 1.27;81.9;49.1]$ attained by NODEs. NODEs shows a much faster error growth outside of the training scenarios. More critically, NODEs fails to capture the sensitivities, resulting in about 100-times larger errors than TRASE-NODEs. Throughout the $u$ range, our proposed TRASE-NODEs consistently tracks both states and sensitivities with much smaller NMSE values. Overall, these results show that explicitly incorporating sensitivity information enables TRASE-NODEs to generalize more effectively from limited training data while providing reliable sensitivity estimates essential for dynamic control applications.


\vspace{5pt}
\noindent \textbf{Test Case 2: Inverter-based Resource (IBR).}
%
IBRs, such as solar PV and wind power plants, interface with the grid through power electronic converters instead of synchronous generators. Their increasing penetration is reshaping the power grid dynamics, as their behaviors are dictated by fast power-electronic controls instead of conventional electromechanical inertia. To represent high-level dynamics, a generic IBR model \cite{deepak2023model} is widely adopted with three key components: plant-level controller (REPC), electrical controller (REEC), and converter model (REGC). In this work, we focus on grid-following IBRs, which behave as controlled current sources. Consequently, the generic model injects active and reactive currents into the grid in response to terminal voltage and frequency. The voltage reference is commonly used as the control set-point for IBRs, to improve their dynamic stability.

For data generation, we use the IEEE 39-bus test system by replacing the synchronous generator in Bus 30 with an IBR model. A 200 MW load increase is applied to induce grid dynamics and the corresponding IBR response is recorded. To ensure consistent initial conditions across different settings, the playback function of the power system simulator is used \cite{foroutan2019generator}. The simulator produces the IBR states $\bbx(t)=[I_d;I_q]$, which are the direct- and quadrature-axis currents. The algebraic variables $\bby(t)=[V_t;f_t]$, namely the terminal voltage and frequency, are fixed by the playback function.  The states are generated for 37 reference values of $u=V_\text{ref}$, in the range of $[1.036,1.045]$. Note that the state $I_q$ is much more sensitive to $V_{\text{ref}}$ \cite{deepak2023model}, as adjusting $I_q$ is more effective in regulating the terminal bus voltage. Without explicit dynamic equations for IBRs, we numerically compute the trajectory sensitivities using finite differences, with the example for $I_q$ given by
\begin{align}
 \displaystyle   s_{Iq}(t) = \frac{I_{q}(t)\big| _{u=V'_{\text{ref}}}-I_{q}(t)\big| _{u=V_{\text{ref}}}}{V'_\text{ref}-V_{\text{ref}}} \label{eqn:s_finite}
\end{align}

\begin{figure}[t]
    \centering
    \begin{tabular}{cc}
    \includegraphics[width=0.225\textwidth]{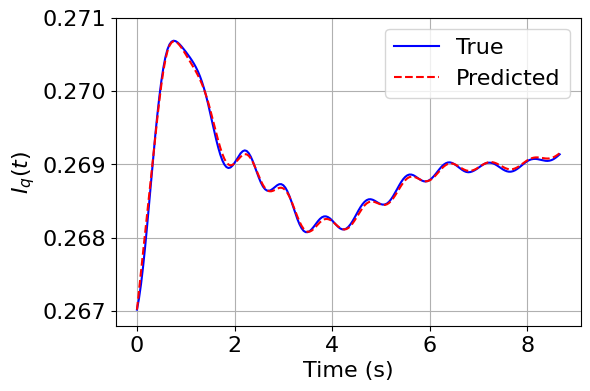} \label{fig:a} &
    \includegraphics[width=0.22\textwidth]{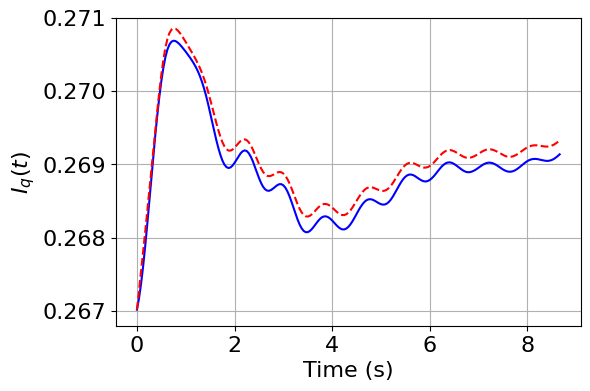} \label{fig:b} \\
    \small (a) & \small (b)\\
    \end{tabular}

  \caption{Example of predicted $I_q$ trajectories at $V_{\text{ref}}=1.0405$ for (a) TRASE-NODEs, and (b) NODEs.}
  \label{fig:Ex2_ExPlot}
\end{figure}

\begin{figure}[t]
    \centering
    \begin{tabular}{cc}
    \includegraphics[width=0.22\textwidth]{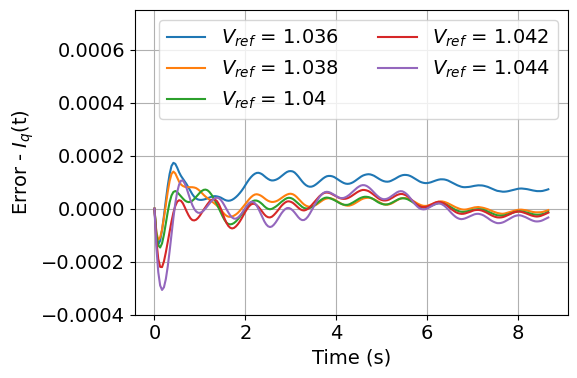} \label{fig:a} &
    \includegraphics[width=0.22\textwidth]{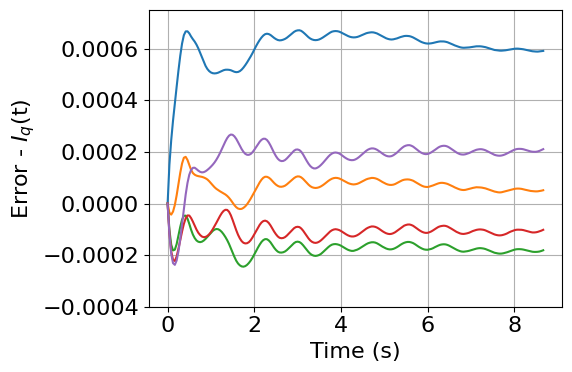} \label{fig:b} \\
    \small (a) & \small (b)\\
    \end{tabular}
  \caption{Normalized $I_q$ prediction error for (a) TRASE-NODEs and (b) NODEs with different $V_{\text{ref}}$ values.}
  \label{fig:Ex2_Error}
  \vspace{-0.5cm}
\end{figure}

For modeling $\bbf_\bbtheta$, we use an NN with five hidden layers consisting each of 32 units and $\mathrm{Tanh}$ activation. The NN input here includes both the state $\bbx(t)$ and algebraic variables $\bby(t)$, along with the scalar control set-point $V_\text{ref}$. For training, TRASE-NODEs use two scenarios of $V_{\text{ref}}=1.03825$ and $1.04275$, with the sensitivity trajectories obtained from~\eqref{eqn:s_finite} using $V'_{\text{ref}}=1.0385$ and $1.043$, respectively. In comparison, NODEs are trained with four scenarios in $V_{\text{ref}}=1.03825$,  $1.0385$, $1.04275$, and $1.043$.

\begin{figure}[t]
    \centering
    \includegraphics[width=0.42\textwidth]{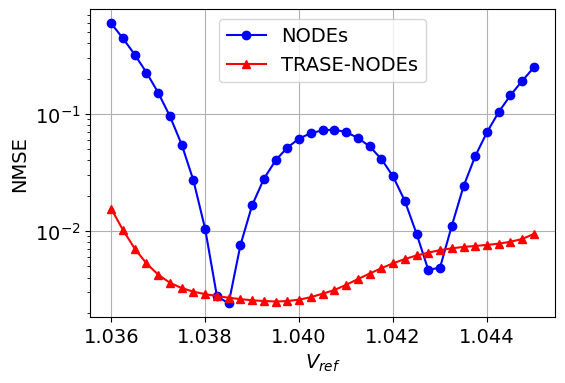}
    \centering
    \caption{Comparison of $I_q$ prediction NMSE for TRASE-NODEs and NODEs over the range of input $V_{\text{ref}}$}
    \label{fig:Ex2_NMSE}
    \vspace{-0.5cm}
\end{figure}

We will mainly present the performance in predicting the state $I_q$, because the other $I_d$ is very weakly dependent on $V_{\text{ref}}$. Performance comparisons of both models have been evaluated across the testing cases of $V_{\text{ref}}\in[1.036,1.045]$, with an example of $V_{\text{ref}}=1.0405$ shown in Fig.~\ref{fig:Ex2_ExPlot}. In addition, Fig.~\ref{fig:Ex2_Error} plots the normalized error in predicting the trajectories for a range of $V_{\text{ref}}$ values. The normalized error is computed as the difference between the true and predicted trajectories divided by the peak value of the true trajectory, allowing errors to be compared across trajectories with different magnitudes. TRASE-NODEs have been observed to produce smaller errors for $I_q$ compared to NODEs. The NMSE across the testing range is shown in Fig.~\ref{fig:Ex2_NMSE}. TRASE-NODEs consistently achieve a lower prediction error compared to standard NODEs and exhibit a smoother error profile across the parameter range. In contrast, the NODEs model shows two pronounced minima that correspond to operating points included in the training dataset. While NODEs achieve low error near these training trajectories, the error increases between them, indicating limited generalization. TRASE-NODEs mitigate this issue by incorporating trajectory sensitivities, which enable the model to better capture how the system dynamics vary with $V_{\text{ref}}$.



%

\section{Conclusions} \label{sec:CON}
This paper presented TRASE-NODEs, a sensitivity-aware extension of NODEs that augment the system jointly with both states and their sensitivities to control inputs. By embedding this information into the adjoint-based training process, TRASE-NODEs can capture input–state dependence that standard NODEs overlook,
achieving better prediction on unseen control set-points. This robustness reduces the need for extensive training data and thereby enhances memory efficiency.
Case studies on a damped oscillator and an IBR demonstrate that TRASE-NODEs consistently achieve lower prediction errors and stronger generalization from limited data than standard NODEs. These findings highlight the value of sensitivity-aware training for dynamic modeling and control, especially in applications where data efficiency and robustness to unseen operation conditions are critical.

\bibliographystyle{IEEEtran}
\bibliography{bibliography}

\end{document}